\documentclass[sigconf, dvipsnames, nonacm]{acmart}

\usepackage{subcaption, graphicx}
\usepackage{algorithm2e}
\usepackage{caption}
\usepackage{epstopdf}
\usepackage{mathtools}
\usepackage{amsmath,amsfonts,amssymb,amsthm}
\usepackage{hyperref}
\usepackage{nameref}
\usepackage{enumerate}
\usepackage{paralist}
\usepackage{tabularx}
\usepackage{float}
\usepackage[cal=cm]{mathalfa}
\usepackage{balance}

\newcommand{\boxscale}{.9}

\newcommand{\figwidth}{.9}

\definecolor{macorchid}{HTML}{7A81FF}
\definecolor{macgrey}{HTML}{929292}
\definecolor{macpurple}{HTML}{663399}
\DeclareMathOperator*{\E}{\mathbb{E}}

\newcommand{\lfm}{Last.fm}
\newcommand{\ml}{MovieLens}
\newcommand{\aminer}{AMiner}

\DeclareMathOperator*{\argmin}{argmin}

\renewcommand\footnotetextcopyrightpermission[1]{} % removes footnote with conference information in first column
\AtBeginDocument{%
  \providecommand\BibTeX{{%
    \normalfont B\kern-0.5em{\scshape i\kern-0.25em b}\kern-0.8em\TeX}}}

\begin{document}

\title{Privacy Shadow: Measuring Node Predictability and Privacy Over Time}

\author{Ivan Brugere}
\email{ibruge2@uic.edu}
\affiliation{%
  \institution{University of Illinois at Chicago}
  \city{Chicago}
  \state{IL}
}

\author{Tanya Berger-Wolf}
\affiliation{%
  \institution{University of Illinois at Chicago\\
    The Ohio State University}
  \city{}
  \state{}}
\email{tanyabw@uic.edu} \email{berger-wolf.1@osu.edu}

\begin{abstract}
The structure of network data enables simple predictive models to leverage local correlations between nodes to  high accuracy on tasks such as attribute and link  prediction. While this is useful for building better user models, it introduces the privacy concern that a user's data may be re-inferred from the network structure, after they leave the application. We propose the \textit{privacy shadow} for measuring how long a user remains predictive from an arbitrary time within the network. Furthermore, we demonstrate that the length of the privacy shadow can be predicted for individual users in three real-world datasets.
\end{abstract}

\maketitle
\thispagestyle{empty}
\section{Introduction}

Networks represent complex relationships which facilitate information retrieval and recommendation due to the local structure of the graph. For example, simple heuristics for link prediction \cite{al2011survey} and attribute prediction \cite{Sen_Namata_Bilgic_Getoor_Galligher_Eliassi-Rad_2008} are extremely effective by exploiting the correlations and homophily in the local network structure \cite{10.5555/2283696.2283849}. Unfortunately, this also means that a node's private attribute or link data can be easily re-inferred from the given graph structure \cite{Zheleva2011, 7576667}.%\footnote{Code, Data, and Reproducibility walkthrough are available anonymously at: https://drive.google.com/drive/folders/1gF7k3NHeA40pf6DSuaff0Q1tGpyS5nVX}

Prior work in the area of network privacy has focused on ensuring a network is ``safe'' to publish under some criteria \cite{10.1145/2020408.2020599, 10.1145/1807167.1807218}, but little work has focused on measuring the privacy dynamics of graph-based predictions over time. Specifically, if a user removes their data from a graph, for how long can we re-infer that data using the last-observed graph structure? We define this length of time as the \textit{privacy shadow}. That is, we don't observe the user directly but leverage the network effects as a secondary image (shadow) of the user over time. 

Network application operators have incentives to build models for users who have left their application or have become inactive, for churn prediction, cold-starting problems, or measuring non-users and inactive users. Therefore, application operators have a strong incentive to build and retain a user's model either directly or \textit{indirectly}. 

Users have the expectation their data is not used after its removal. In fact, new regulations such as the EU General Data Protections Regulation (GDPR) explicitly require transparent removal of user data \cite{eu-269-2014}. For a network operator this means not only ensuring the removal of an individual user's data, but also secondary data representations, and their encoding within subsequent machine learning model representations. This transparency presents a unique challenge, especially in end-to-end machine learning models where data representation and model representation are tightly coupled. 

In this work, we examine the relationship between the privacy of individual user data and secondary (especially network) representations and predictive models, particularly over time. In order to focus on our methodology, we use simple models and graph representations. However, our formulation is general and can be applied to more sophisticated predictive models, data representations such as node embeddings, or end-to-end graph neural network models.

For the purposes of this work, a representation satisfies user privacy when the model trained on the representation are no better than a \textit{zero-information} model for predicting the target data of a user (e.g. sensitive attributes, attribute distribution, etc). This definition also extends over time, when a prior representation is no longer predictive at a future time-step with respect to a zero-information model.

\subsection{Contributions}

In this work, we propose and measure the concept of the \textbf{privacy shadow} in dynamic networks. This is a measure of a node's attribute distribution predictability over time, from the last observed network instance. Specifically, we do the following:
\begin{itemize}
    \item Define the privacy shadow for node-attribute prediction tasks.
    \item Provide interpretable, baseline criteria for evaluating a node's predictability over time.
    \item Demonstrate this measure on three real-world datasets with distinct privacy shadow profiles.
\end{itemize}

\section{Related Work}
We organize our work around two primary areas (1) dynamic network inference and prediction and (2) data privacy in networks. 

Prior work in the study of networks has shown that simple heuristics for tasks such as link prediction \cite{al2011survey} and attribute inference \cite{Sen_Namata_Bilgic_Getoor_Galligher_Eliassi-Rad_2008} are very effective by leveraging the complex structure and homophily of networks of networks. \cite{10.5555/2283696.2283849}. 

Similar work also exploits this local task performance to infer and evaluate \textit{non-spurious} networks from underlying non-network data \cite{brugere2018network, 10.1145/3184558.3191525}. In a similar direction, prior work define an efficient summarization of a \textit{given} network \cite{doi:10.1137/1.9781611973440.11} often using Minimum Description Length approaches \cite{10.1002/0471667196.ess1641.pub2}. Both areas of work measures the predictive utility of some network definition. In this current work, we do this across time, against non-network predictive baselines.

Prior work in the area of network privacy \cite{Zheleva2011} has focused on ensuring a network is ``safe'' with guarantees for attacks for de-anonymization, inferring node degree, attribute or edge values \cite{10.1145/2020408.2020599, 10.1145/1807167.1807218, 10.1371/journal.pone.0130693, 8094235}. Often this work measures the \textit{utility} of the sanitized representation for downstream tasks, for example, recommendation\cite{Meng2019}. While many aspects of privacy are studied with respect to static networks, very few prior works consider dynamics \cite{rossi2015k, tai2011identities, 10.5555/1863190.1863196}.  

Extensive work focuses on measuring domain-specific inference attacks of sensitive attributes, including physical location and mobility \cite{alessandretti2018evidence}, demographics including gender and age \cite{dong2014inferring}. Other work estimates tie strength between individuals and demographic or other node-attribute groups \cite{10.1007/978-3-319-50011-9_26, 10.1145/1772690.1772790, 10.1145/3154793}. 

Our work differs from these two areas in that we use predictability over time as a general measure of the dynamics of individuals and the population, rather than mitigating specific sensitive-data attacks or sanitizing data for publication.

\begin{figure*}[t]
  \centering
  \includegraphics[width=0.75\textwidth]{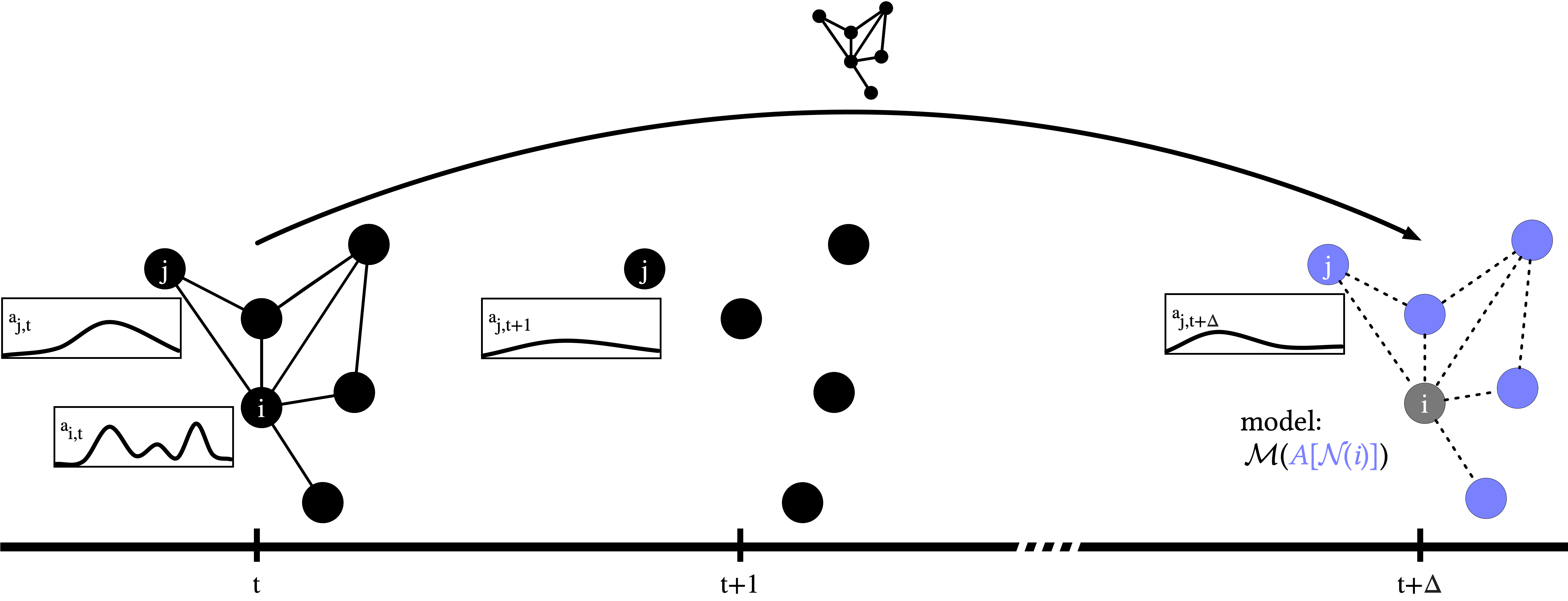}
  \caption{A schematic of the network prediction \textbf{interval}. We observe the local neighborhood of node $i$ at time $t$. Attribute distributions are represented by the call-out. Node $i$ is removed from the graph at time $t+1$. We then observe only the changing attribute distributions of the remaining nodes. At time $t + \Delta$ we receive the graph structure observed at time $t$, and construct a model $\mathcal{M}$ on the \textit{current} attributes in the \textcolor{macorchid}{\textbf{neighborhood}} of $i$ (excluding $i$). The dashed edges indicate this may not be the \textit{current} graph structure. Node i is the \textcolor{macgrey}{\textbf{target}} for label inference.}
  \label{fig:schematic}
\end{figure*}

\section{Methods}

\subsection{Notation}

This work measures predictive decay over time of nodes in attributed, dynamic graphs.

Let $G = (V, A, E)$ represent be an attributed graph with nodes $v_i \in V$, edges $e_{i,j} \in E$, and node-attributes $a_i \in A$. For notational convenience $G_{[1...T]}$ refers to a time-series of graphs, where $G_t$ is the graph at time step $t$, $1 \leq t \leq T$. Similarly, $a_{i,t}$ denotes the attributes of node $i$ at time $t$. 

We define a node neighborhood function on node $i$: 

\begin{equation}
\mathcal{N}(E,i) \rightarrow S, \textrm{where } S \subseteq V
\end{equation}, 

This need not be graph adjacency but any local node sampling function on $E$. Below, we omit $E$ from the notation for simplicity.

For further notational convenience, let $L \subseteq A$ be predictive \textit{targets}. A local \textit{model} predicts the output label(s) of node $i$ as a function of the attributes in the neighborhood: 

\begin{equation}
\mathcal{M}(A[\mathcal{N}(i)]) \rightarrow \hat{l}_i
\label{eq:model_general}
\end{equation}

\subsection{Network prediction horizon}

We formulate the network horizon prediction problem as re-inference of node $i$'s labels at time $t + \Delta$ using the graph fixed at time $t$:

\begin{equation}
\mathcal{M}(A_{t + \Delta}[\mathcal{N}(E_t,i)]) \rightarrow \hat{l}_{i, t+\Delta}
\label{eq:model_filled}
\end{equation}

In Equation \ref{eq:model_filled}, the model has access to the neighborhood \textit{attributes} at time $t + \Delta$, but the graph from time $t$. This simulates the user leaving the system at time $t$ and no longer generating attribute data. We retain the network at the time of leaving. We are unable to access the node's attributes $a_{i, t+}$, but only the attributes of $i$'s \textit{last known} neighboring nodes still in the system.

Our methodology is agnostic to the definition of the local model $\mathcal{M}()$. Future work will focus on model development to improve re-inference accuracy, but the methodology is just as applicable on better models. For now, for clarity we'll focus only on simple aggregation functions (e.g. attribute mean, neighbor-attribute sampling) under a known attribute-to-label mapping.

% We purpose a task of re-inferring node attributes at some time horizon, omitting the node's own data. We retain some small meta-data of the node, and some number of the node's incident edges.

% Let time $T_i$ be a time where node $i$ is removed from the system. Let $\mathcal{M}$ be some function or function-set to represent nodes:

% \begin{equation}
% \mathcal{M}(a_i) \rightarrow m_i
% \end{equation}

% where $m_i \in M$ is a model of node $i$. For some fixed predictor function $\mathcal{P}$ on all $M$: 

% \begin{equation}
%     \mathcal{P}(M, T + \delta) \rightarrow \hat{A}
% \end{equation}

% Finally, we can formulate a complete problem statement:

% \begin{equation}
%     L(A, \hat{A}, M) = |A - \hat{A}| \times \alpha \mathrm{log}(\mathrm{bytes}(M))
% \end{equation}

% Therefore we have the following problem formulation:

% \begin{algorithm2e}%[H]
%   \SetAlgorithmName{Problem}
%   \\
%   \\
%   \KwGiven{Attributed graph $G= (V,A,E)$, node model function-set $\mathcal{M}$, predictor \mathcal{P}}
%   \KwFind{Node models $M^*$}
%   \KwWhere{...}
%   \caption{...}
%   \label{p:networkinference}
% \end{algorithm2e}

\subsection{Proposed measures}

We organize our methodology by network prediction \textbf{intervals}. Figure \ref{fig:schematic} gives a high-level overview of a prediction interval. At time $t$, we observe a graph and associated attributes. For simplicity, we only visualize a target node $i$ and its neighborhood, and example attribute distributions. From time $t+1$, both the graph structure and the target node $i$ are unobservable. At some time $t+\Delta$, we use a local model $\mathcal{M}()$ to re-infer the target labels of node $i$ (\textcolor{macgrey}{grey}), using the graph structure (\textcolor{macorchid}{blue}) we previously observed at time $t$.

A prediction \textit{interval} measures the model's re-inference performance at each step $\Delta=[0,1,2,...t_{max}]$, where we consider $t_{max}$ the end of available data. Recall our model (Equation \ref{eq:model_filled}) produces a predicted label distribution of node $i$ at time $t$. Then an interval is simply the sequence of the model outputs starting at time $t$:

\begin{equation}
r_{i,t} = [\hat{l}_{i, t+\Delta}] : \forall \Delta \in [0,1,...,t_{max}] 
\end{equation}

\begin{figure}
  \centering
  \includegraphics[width=0.95\columnwidth]{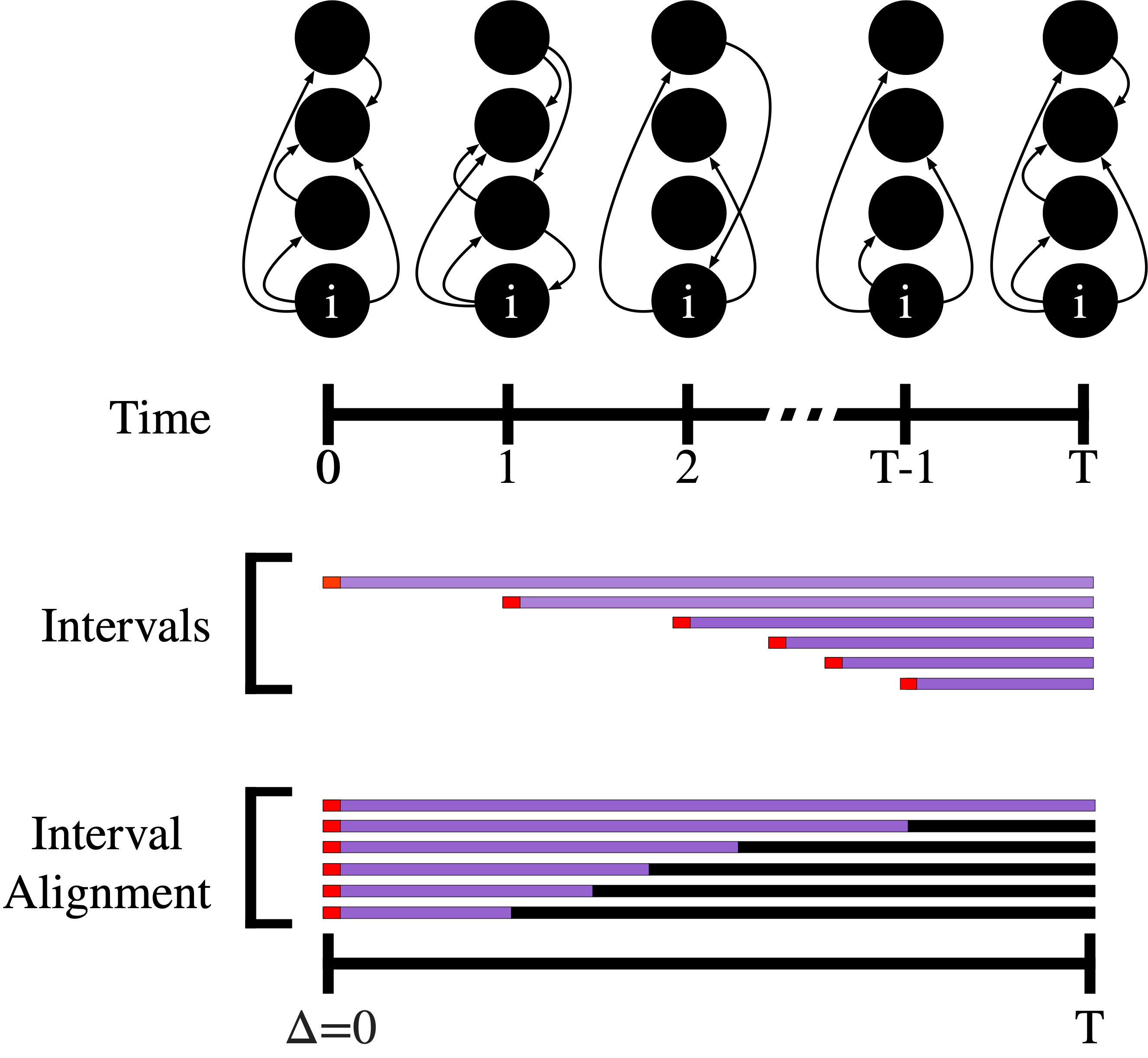}
  \caption{An overview of interval alignment. (Top) shows a toy neighborhood network of $i$ at each time step. (Middle) a \textit{interval} is measured on the graph starting at time $t$ (\textcolor{red}{red}), and measures the predictability of node $i$ using this fixed network at each available time step $>t$ (\textcolor{macpurple}{purple}). Finally, (Bottom) we align the intervals at their starting point and measures expectations on a particular time shift $\Delta$ of this alignment (e.g. across red values), where black indicates null fill-values.}
  \label{fig:intervals}
\end{figure}

This is a sequence of length $t_{max}+1$. It is important that two overlapping interval segments, e.g. $r_{i,0}[1]$ and $r_{i,1}[0]$ are not identical.\footnote{We use bracket notation to denote a particular element in this sequence: e.g. $r_{i,t}[j]$.} In this example, both make predictions for node $i$ at time $t=1$. However, the graph topology used by the predictive model changes between these two intervals. Figure \ref{fig:intervals} illustrates this intuition. At the top, we see the time series of network structures over time, on a toy neighborhood of node $i$. Below that we see a collection of intervals with different starting points (in \textcolor{red}{red}), until the end of available data.  

Next, we align intervals by their starting points to measure node $i$'s predictability for each $\Delta$ \textit{in expectation} over starting-points. In Figure \ref{fig:intervals} this corresponds to expectation over each column in the alignment:

\begin{equation}
    \mathcal{R}(i) = \E(\mathcal{L}(r_{i}[\Delta])) : \forall \Delta \in [0...T]
\end{equation}

where $\mathcal{L}$ is some loss or accuracy measure. This is a sequence measuring node $i$'s expected predictability after $\Delta$ steps from \textit{every} prior observed network.

For notational convenience, we also define this measure over a set of nodes $S\subseteq V$ on a fixed time $\Delta=t$:

\begin{equation}
    \mathcal{R}(S, t) = \{R(i)[t] : \forall i\in S\}
    \label{eq:rs_exp}
\end{equation}

In Figure \ref{fig:intervals} for each node in S, we can take the expectation over the red column at $\Delta=0$. This yields the distribution of predictability of nodes at a fixed $\Delta$. This is a key distribution for the remainder of our work and will define a \textit{reference distribution} to measure nodes against.

% vector with the t-th position measuring the expected predictability of node $i$ at t$\Delta$ where we align intervals at $\Delta=0$ and vary the \textit{starting} time $t$ of each interval, appending null values onto shorter intervals to yield fixed T-length intervals. 

\subsection{Measuring the Privacy Shadow}

%For a given node, we aim to measure its \textit{rate} of decay in predictability, using the observed graph at time-shift $\Delta=0$. 

%At a fixed $\Delta=t$, we defined the distribution of predictability across nodes: $\mathcal{R}(S, t)$. We measure the rank-statistic (e.g. percentile) of node $i$ within this distribution:

Using Equation \ref{eq:rs_exp} we can measure the decline in predictability of the node $i$ using models trained up to the time $t_{ref}$. We do this with a rank-statistics (e.g. percentile) with respect to a reference distribution:

\begin{equation}
    \mathrm{rank}_i(t|t_{ref}) = \mathrm{rank}(\mathcal{R}(V, t_{ref}), \mathcal{R}(i)[t])
    \label{eq:rank}
\end{equation}

Concretely, this is node $i$'s percentile within the reference distribution at some arbitrary time $t$. When $t_{ref} = t$, this is the ranking statistic of node $i$'s initial percentile within the distribution, i.e. at $\Delta=0$. When $t > t_{ref}$, we measure node $i$'s predictability at a later time, but still within the \textit{reference distribution}. 

We define the \textit{$\eta$-shadow} of node $i$ as the minimum $\Delta$ such that \textit{all} subsequent rank measures are less than $\eta$:

\begin{equation}
    \mathrm{rankshadow}_i(t_{ref},\eta)=\mathrm{\argmin}_{t} \mathrm{rank}_i(t_{ref}, t_{2}) \leq \eta, \forall t_{2} \geq t 
    \label{eq:shadow}
\end{equation}

The rank shadow is a measurement of \textit{time}. To interpret this measure, assume node $i$ satisfies the $10$-percentile privacy shadow with respect to the $\Delta=0$ distribution, at index $t=15$. This means that after $t \geq 15$ time-steps, the node's prediction ranking in expectation is in the bottom $10$ percentile of the reference distribution at $t_{ref}=0$.

We choose a rank-based measure because it is adaptive to reference distribution and serves as a non-parametric normalization. Depending on application, it may be appropriate to measure the nodes with respect to (1) an absolute prediction threshold, or (2) a parametric reference distribution rather than the empirical $\Delta=0$ distribution. Below, we fix the $\eta$ threshold with respect to a zero-information, non-network baseline. Network sampling methods may also be appropriate, e.g. measuring against a reference distribution using community/cluster sampling. However, in our definition of privacy, these would not serve as a zero-information baselines.

% Figure \ref{fig:bounded_lastfm} gives further intuition of this measure on the \lfm{} dataset we later measure in detail. There are many nodes with a small index ($ \leq 10$). This shows that low-ranked nodes with respect to the reference distribution remain low-ranked over the varying $\Delta$. Furthermore, approximately $52\%$ of nodes never achieve a $10$-percentile bound. 

% Either (1) the node is not predictable at any time-shift and is bounded at $\Delta=0$, or (2) it quickly decays from a higher ranking. If the returned

\section{Evaluation Details}

\subsection{Neighborhood and Graph Models}

\subsubsection{Graph representation}
In this work, we construct a training network with some $w$ time steps. For induced $k$-NN networks, we simply calculate $k$-NN on the aggregated attribute distribution over $w$, for the annual co-author graph, we simply union over all edges in $w$.We treat these $w$ time steps as an aggregated time step at $\Delta=0$ of a prediction interval. 

\subsubsection{Neighborhood Model}

In this work, we apply the simple \textit{neighborhood average} as the model $\mathcal{M}()$ for aggregating node attributes over the graph. More sophisticated models producing attribute/label distributions or dense embeddings (e.g. using Graph Neural Networks) could be similarly measured. We choose this simple model to focus our attention on the methodology for measuring the privacy shadow. We use bootstrap sampling with $p=0.5$ drawn from the network neighborhood, so the effective neighborhood size is half that of our induced $k$-NN graph, measured over sample realizations. 

To produce the predicted labels from the aggregated attribute distribution, we rank label support using the attribute-label mapping and measure Jaccard similarity on the same number of ranked and true labels.

\subsection{Non-network Baselines}
\label{subsec:baseline}
We measure the privacy shadow using an observed network. However, there are several non-network baselines which require zero information about an individual node.

Recall that we defined the privacy shadow with respect to some predictability threshold $\eta$. We use the expectation of the following baseline distributions as criteria for when a node is no longer predictable. We use a straightforward population sampling, attribute sampling, and true random sampling. Concretely, the length of the privacy shadow can be interpreted as the time until the node's prior network neighborhood is no more predictive than random population sampling on current data.

\subsubsection{Population sampling}

For each node, we sample ($k=20$) random nodes and aggregate attributes by the neighborhood model as in the network model. We sample these independently at each step in the node's prediction interval. We denote this baseline by `population sampling` (\texttt{pop}).

\subsubsection{Attribute sampling}

For each node, we sample the \textit{median} attribute density (e.g. number of non-zero attributes) according to the empirical likelihood of unique attributes over all nodes. These are samples of the `average' node with respect to both unique attribute likelihood and node-attribute density (e.g. user activity). We denote this baseline by attribute sampling (\texttt{attr}).

\subsubsection{Random attribute sampling}

While the above baselines use random sampling, they should \textit{not} be considered random models. Instead they are \textit{global}, non-network models i.e. ``no network.'' For a truly random model, for each node we uniformly and independently sample from the set of all labels. We denote this baseline as random attribute sampling (\texttt{rand}).

In practice, we set $\eta$ in the privacy shadow definition (Eq. \ref{eq:shadow}) as the \textit{maximum} of all baseline expectations.

\subsection{Measuring node cohorts}

We define a $s$-length \textit{complete} interval as an interval having non-null predictions for all $\Delta=[0...s]$. The number of nodes satisfying this definition is non-increasing for larger $s$. Without accounting for this, measuring distributions of node statistics over varying $\Delta$ would produce incomparable results because the measured population is changing per $\Delta$ timestep. 

We therefore build a cohort with $s$-length \textit{complete} intervals and measure all further statistics on these populations:

\begin{equation}
\mathcal{R}^s(V) = \{\mathcal{R}(i) \textrm{ is } s\textrm{-length complete}, \forall i\in V\}
\end{equation}

\section{Datasets}

Our work is primarily applied to dynamic, attributed, graph-structured datasets with high attribute and label cardinality. Tables \ref{tab:data1}-\ref{tab:data2} give an overview of datasets used in this work.

\subsection{Last.fm}

Last.fm is a music-focused social network where users can programmatically log their listening history, and interact with other users. In this application, users are represented by nodes. Users have time-stamped song plays, and user-attributes are the play distribution of artist within a time window. Edges represent either (1) friendship in the social graph or (2) an induced graph using user-attribute similarity.

Prior work collected the complete listening data of more than 1M users between 2005 and 2016\cite{mlg2017_13}. Similar to prior work, we use the connected subgraph of 20K users discovered from crawling the social graph from a seed node.

We collect a new label-set index, containing the top-500 most frequent artist-tags. For each tag, we collect the top-1000 most tagged artists. Within a time-step, a node is labeled with the tag-label if the user has listened to more than ($k=5$) of these top 1000 artists within the interval. We split the timeline into $50$ equal-length intervals, yielding $85$ days per time-step.

We induce a $k$-nearest neighbor graph ($k$=20) from node-attribute cosine similarity. We experimentally vary $k$ and verify our privacy shadow measurement is consistent, but the precise tuning of this parameter is not important to demonstrate our methodology.

\subsection{Movielens}

Movielens is a platform for users to rate movies and organize movies with meta-data tags. We use the Movielens 25M movie rating dataset \cite{Harper:2015:MDH:2866565.2827872} with data from 1997 to 2019, yielding 162K users (i.e. nodes) with 25M ratings (i.e. node-attributes). Similar to our organization of the Last.fm dataset, we group the $1000$ most common movie-tags as our target label-set $L$. We create $50$ equal-width intervals over the time duration of the dataset, yielding $177$ days per time step.  

 \begin{table}[h]
 	\centering
 	\resizebox{\boxscale\columnwidth}{!}{
	\begin{tabular}{|c|c|c|c|c|}
		\hline
		Dataset & Nodes & Edges & Attributes & Labels \\\hline
		\lfm{} 20K \cite{mlg2017_13} & Users & K-NN & Artist listens &  User tags\\
		\ml{} (25M) \cite{Harper:2015:MDH:2866565.2827872} & Users & K-NN & Movie ratings & User tags\\ 
		\aminer{} (v2) \cite{10.1145/2740908.2742839, tang2008arnetminer} & Authors & Co-authorship & Paper keywords & Paper keywords\\
		\hline
	\end{tabular}}
 	\caption{Datasets} 
 	\label{tab:data1}
 \end{table}

 \begin{table}[h]
 	\centering
 	\resizebox{\boxscale\columnwidth}{!}{
	\begin{tabular}{|c|c|c|c|c|}
		\hline
		Dataset & $|V|$ & Median degree & $|A|$ & Unique $|L|$ \\\hline
		\lfm{} 20K \cite{mlg2017_13} & 19,990 & 20 & 1.2B & 498\\
		\ml{} 25M \cite{Harper:2015:MDH:2866565.2827872} & 162,542 & 20 & 25M & 970\\ 
		\aminer{} v2 \cite{10.1145/2740908.2742839, tang2008arnetminer} & 1,322,078 & 3 & 136B & 634K\\
		\hline
	\end{tabular}}
 	\caption{Summary statistics of datasets} 
 	\label{tab:data2}
 \end{table}

\subsection{Open Academic Graph 2019}

The Open Academic Graph  \cite{10.1145/2740908.2742839, tang2008arnetminer} dataset combines data from Microsoft Academic Search and ArnetMiner. We use the ArnetMiner dataset between the years of $1990$ and $2019$. We use yearly time-steps since this is the smallest granularity of paper publication in the dataset. This dataset contains $1.3$M authors (i.e. nodes) with at least $20$ total publications. Node-attributes are given by the aggregated publication keywords within each year, and edges are defined as the top-$200$ co-authors within that year. 

Unlike Last.fm and Movielens, our prediction target is the keyword distribution itself rather than a reduced label-set. The edge definition is directly from data rather than induced, yielding a variable degree distribution with median degree of $3$.

\section{Results}
\label{sec:results}

\subsection{Measuring the Privacy Shadow}

In this section, we walk through the methodology for measuring the privacy shadow on the \lfm{} dataset and present other datasets in Section \ref{sec:discussion}.

\begin{figure}
\centering
  \includegraphics[width=\figwidth\columnwidth]{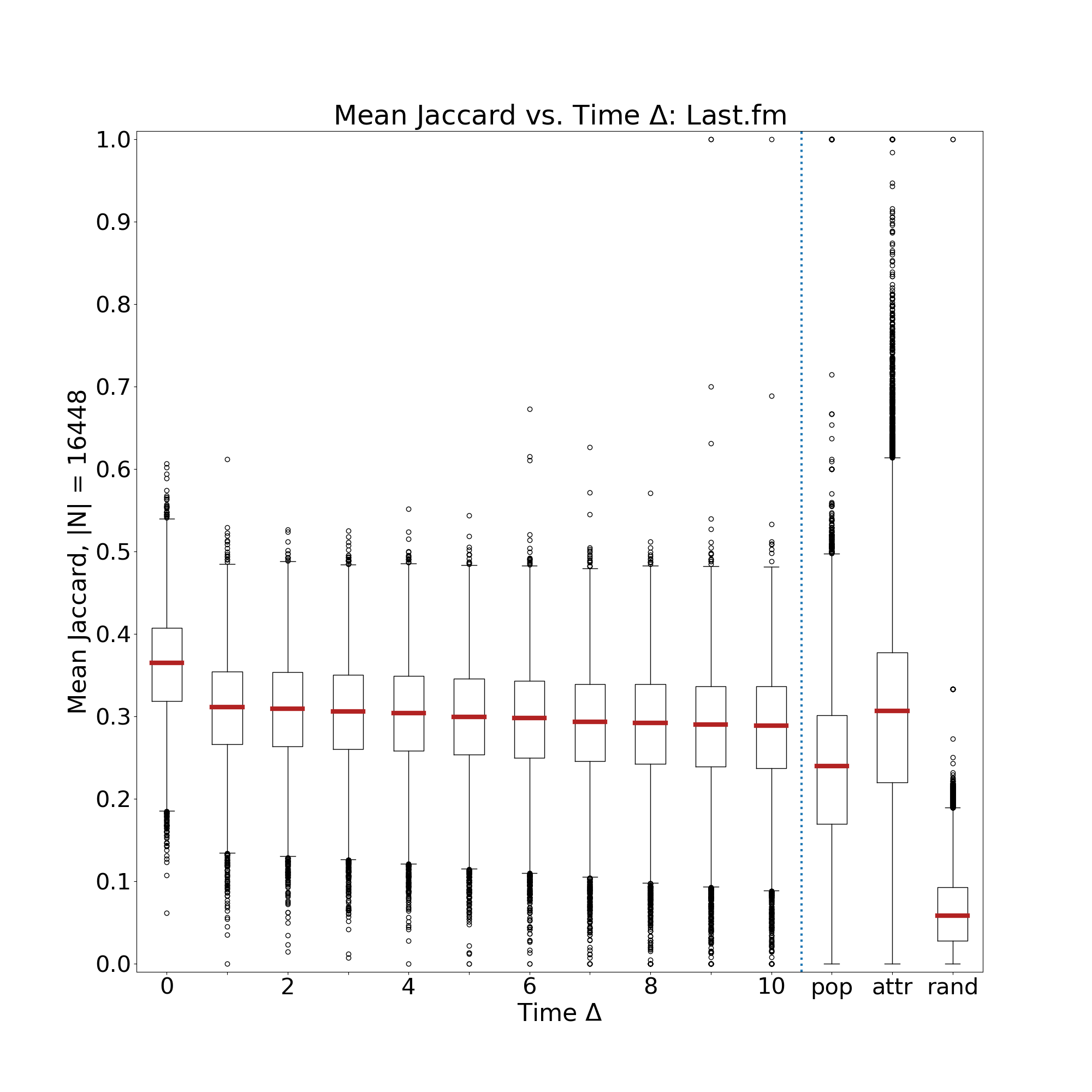}
  \caption{Predictive performance on \lfm{} over $\Delta$ time shifts, for the sub-population with $s$=10 complete intervals ($|N|$=16407). A data-point within each box-plot is a node's mean Jaccard similarity between predicted and actual labels at an $x = \Delta$, over all of the node's prediction intervals. The non-network baselines, \texttt{pop}, \texttt{attr}, and \texttt{rand} are reported on the right.} 
  \label{fig:deltas_lastfm}
\end{figure}

Figure \ref{fig:deltas_lastfm} shows the node prediction accuracy in terms of Jaccard similarity between predicted and real labels on the \lfm{} dataset. We measure the cohort which are at least $s=10$ complete ($\approx 2$ years of data), yielding $|N| = 14712$ measured nodes. In each box-plot, a data-point is the mean Jaccard similarity of an individual node at a $\Delta$ timestep aggregated over all its prediction intervals. A node may have a varying number of evaluated intervals, depending on the node's available data. $\Delta=0$ measures the autocorrelation in re-inferring labels from the same node attributes which induced the $k$-NN network, under mean neighborhood aggregation. 

On the right side of Figure \ref{fig:deltas_lastfm}, we evaluate the same set of nodes, under varying non-network baselines (Section \ref{subsec:baseline}). Attribute sampling (\texttt{attr}) performs approximately as well in expectation as the network model at $\Delta=3$, although the baseline has much higher variance. The random label sampling $\texttt{rand}$ performs much worse than all other methods. This illustrates that random population sampling is \textit{not} a random labeling i.e. a purely random method.

\begin{figure}
\centering
  \includegraphics[width=\figwidth\columnwidth]{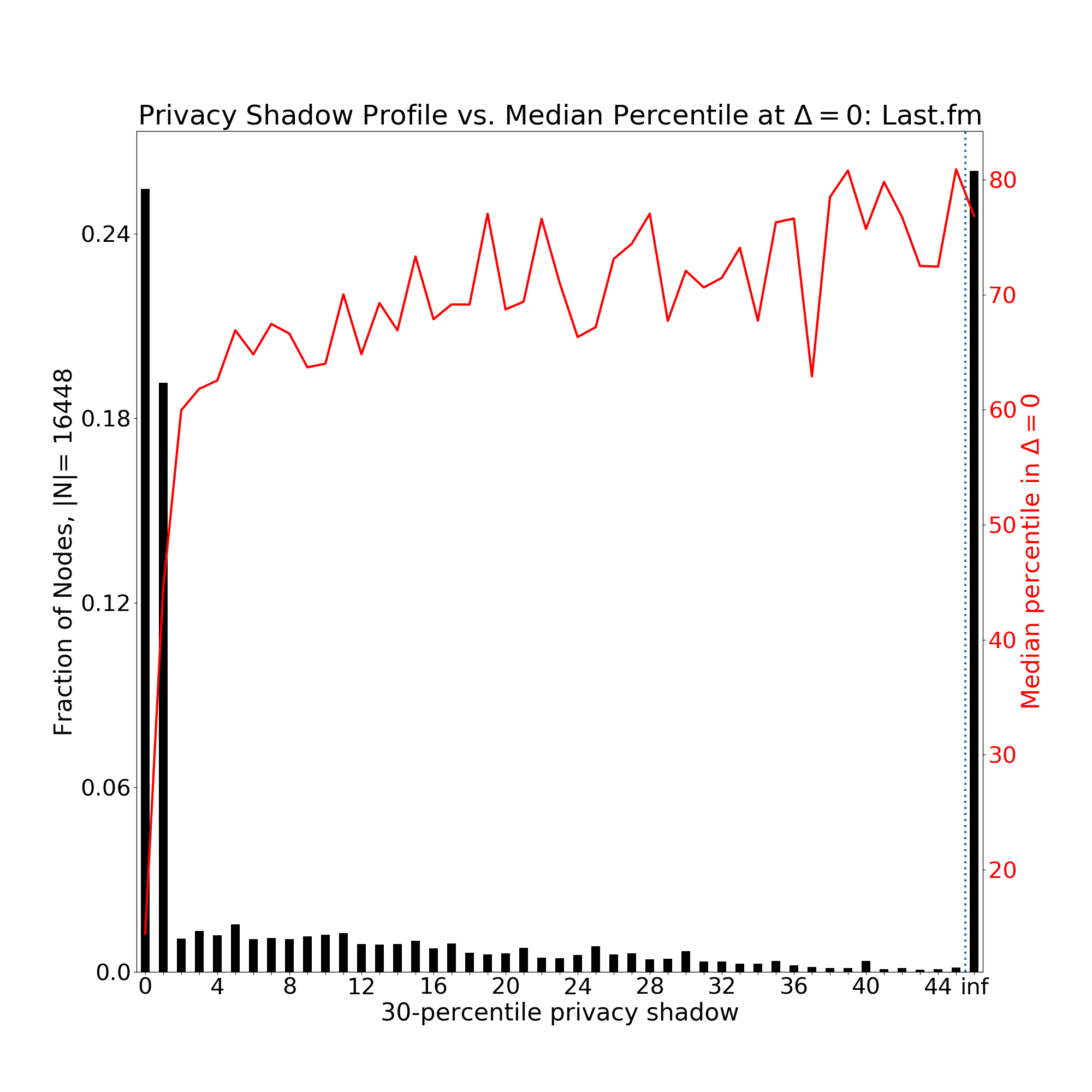}
  \caption{The Privacy Shadow profile on \lfm{}. The $x$-axis correspond to nodes with $x=t$ privacy shadow (Equation \ref{eq:shadow}). $\eta=30$-percentile is set against the reference distribution ($\Delta=0$) by the maximum performing baseline (\texttt{attr}). The $y$-axis reports the normalized fraction of nodes at this privacy shadow length.  \textit{Take-away:} more predictive nodes at the start of an interval ($\Delta=0$) are predictive for longer. $> 25\%$ of nodes are indefinitely predictive.} 
  \label{fig:bounded_lastfm}
\end{figure}

Figure \ref{fig:bounded_lastfm} shows the privacy shadow of the same set of $s=10$ complete nodes on \lfm{} ($|N| = 14712$). The $x$-axis caption reports the $\eta$ threshold given by the median of the \texttt{attr} baseline distribution, which corresponds to the $31$st percentile of the $\Delta=0$ reference distribution. This means that the bottom 31\% of nodes perform no better than sampling from the empirical attribute distribution. The privacy shadow is the minimum time steps until this threshold is an upper bound to a node's performance for \textit{all} subsequent time steps. For example, $24\%$ of nodes satisfy the $31$-percentile threshold bound at $\Delta=0$. It is not necessarily true that $0.31$ have a privacy shadow length of $0$, because nodes can stochastically improve performance at later steps, yielding a non-zero privacy shadow. Nodes that never satisfy the bound are indicated by \texttt{inf} on the right of the figure. Although this cohort consists of nodes with $s$-complete intervals, we measure their \textit{individual} privacy shadow over all available data, so nodes can have a privacy shadow greater than $10$ time-steps. In Figure \ref{fig:bounded_lastfm}, the red $y$-axis reports the median percentile of nodes in this bin, with respect to the reference distribution. For example, nodes with infinite privacy shadow are in expectation at approximately the $75$th percentile in the reference distribution.

There are two key observations in these figures. First, for over half of nodes, the network model is either \textit{never} predictable or is \textit{indefinitely} predictable with only a small number of neighbors. Concretely, for predictive nodes this means the data of $10$ neighbors performs better than attribute sampling, even when those neighbors were chosen as much as $45$ time steps (10 \textit{years}) ago. Second, the length of the privacy shadow is positively correlated with the node's percentile in the reference distribution. This simply means that nodes which are initially predictive remain predictive for longer. 

\begin{figure}
\centering
  \includegraphics[width=\figwidth\columnwidth]{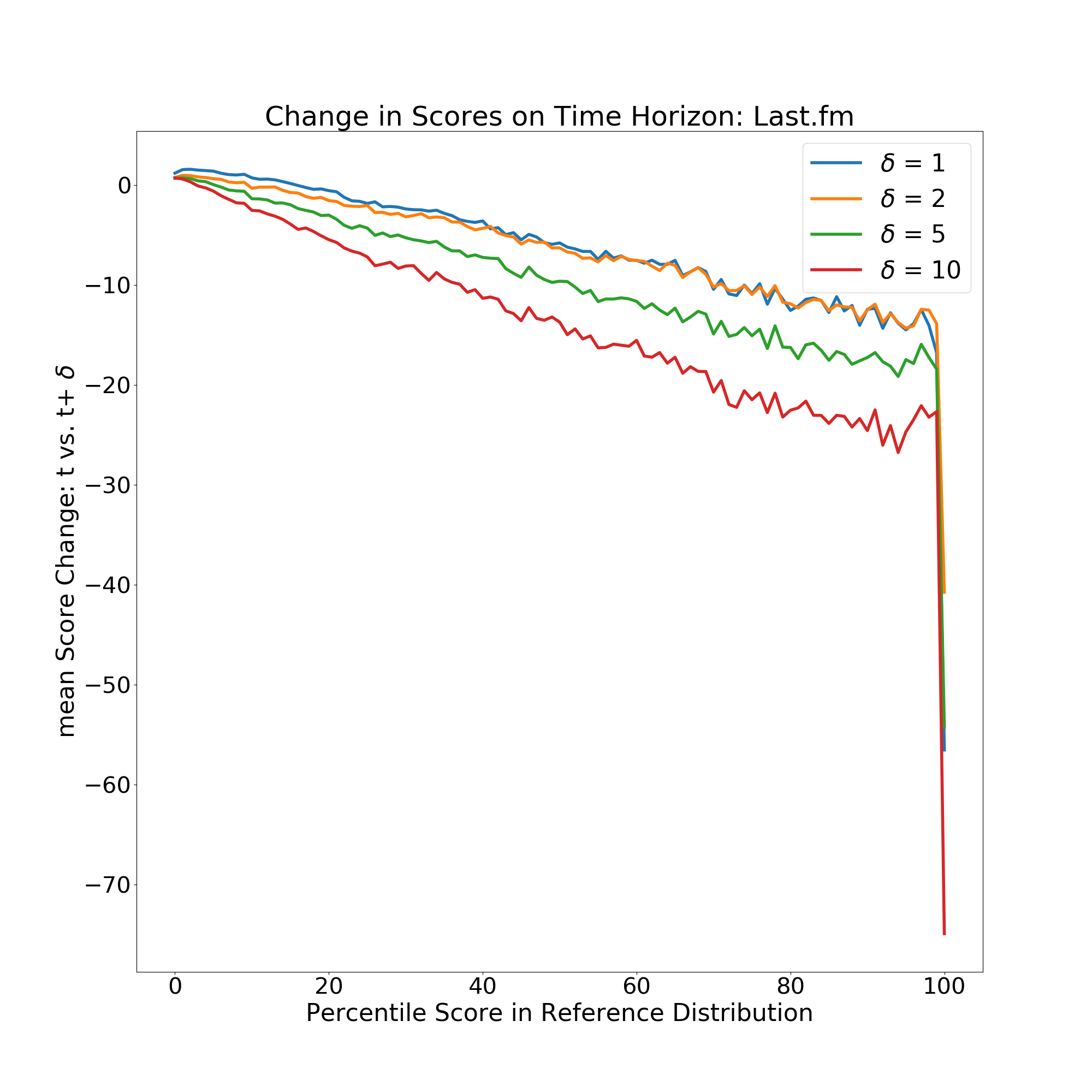}
  \caption{\lfm{} score vs. score on a $\delta$ time horizon. For the expected predictability of a node over its prediction intervals, reported with respect to its percentile in the reference distribution (x-axis). We group such observations by integer percentile and report the mean change in score (y-axis). This shows that higher ranked nodes have higher negative change. } 
  \label{fig:lastfm_meandiff}
\end{figure}

Figure \ref{fig:lastfm_meandiff} summarizes the difference in a node's predictability at an arbitrary time step. For any time step $t$, the $x$-axis is the node's rank at time $t$ with respect to the reference distribution. We group observations by discrete percentiles. The y-axis shows the mean \textit{difference} at time t and time $t+\delta$ for nodes at this percentile. 

This show that for \lfm{}, the degradation in rank is not only correlated with the initial rank of the node, but the loss in predictability scales linearly for any current observation of the node. These values have a lower bound of the diagonal $y=-x$. At $\delta=10$, the nodes measured in \lfm{} are approximately $y=-.25x$.

\subsection{Predicting the Privacy Shadow}
\label{subsec:predicting_privacy}

We now aim to achieve our initial goal of predicting an \textit{individual} node's privacy shadow. In contrast to simply measuring the privacy shadow post-hoc, we aim to give a user some guidance about their predictability if they remove their data at the current time. We demonstrate the utility of predicting from rank score trajectories (Equation \ref{eq:rank}) over varying time-steps, generated from our methodology above. 

\begin{figure}
\centering
  \includegraphics[width=\figwidth\columnwidth]{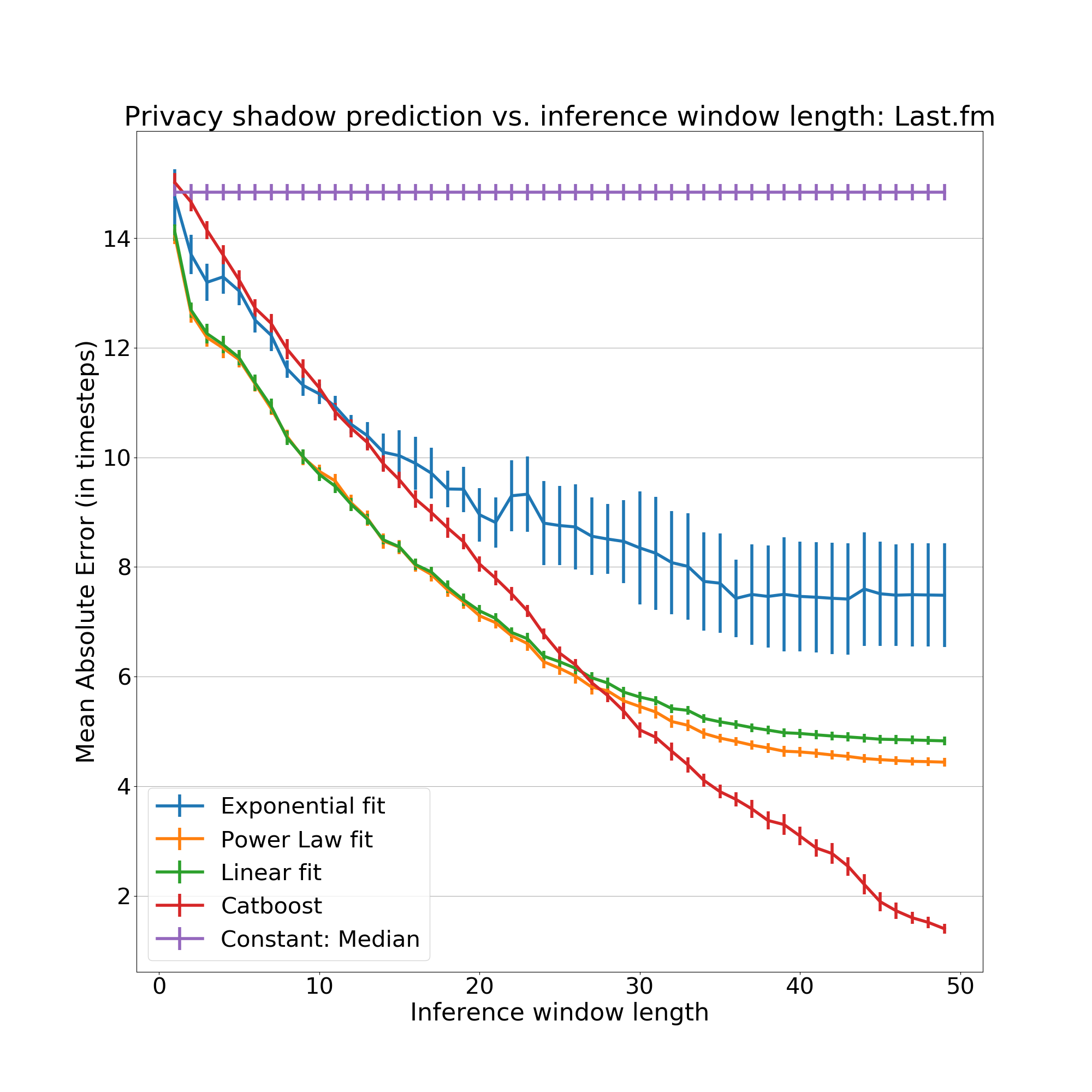}
  \caption{Prediction of privacy window length on \lfm{}. Reports the mean absolute error (MAE) of a model trained on full prediction trejectories, and (x-axis) varying the length of data visible to the model in inference by yielding null data for $t > x$. A lower MAE shows the model better predicts the privacy window length given longer score trajectories.} 
  \label{fig:lastfm_shadow_preds}
\end{figure}

In this problem setting, we train a model offline on the full rank trajectories of a training set of users. We simulate users leaving the system by varying available data given to a simple gradient boosting model\footnote{https://catboost.ai/} using null fill values after varying time steps. Figure \ref{fig:lastfm_shadow_preds} reports the mean absolute error (MAE) of predicting the length of privacy shadow from only rank trajectories, after observing an increasing number of steps from the node (x-axis). We train models over (|N|=30) randomly re-sampled train-test splits. This MAE estimates the variance of our model for a particular user with a given number of steps of observed data at opt-out. For example, with $20$ steps of rank trajectories ($4.5$ years), the model predicts with MAE of $6$ time steps ($1.25$ years). This can be greatly improved in future work by considering better sequential models, or building an auto-regressive or similar model directly from attributes.

We use a similar procedure to infer parameters of a line-fitting model for extrapolation. For each training node's complete rank trajectory, we fit three models: Power Law ($y=ax^\lambda +c$), Exponential ($y=e^{\lambda x} +c$) and Linear ($y=ax+b$). At test time, we vary the observed score trajectory length as above. For each sub-trajectory, we select the fit parameters of the $1$-nearest neighbor of training instances. We then extrapolate on this model and return the predicted privacy shadow length.

This method gives us more information about the model of dynamics for each dataset. While both Power Law and Linear produce a good fit, the Power Law exponents tend to be yield a near-linear model with small exponent: $y = -14x^{0.42}$ at median exponent. 

\subsection{Discussion}
\label{sec:discussion}

\begin{figure*}[ht]
\centering
\begin{subfigure}{.33\textwidth}
  \includegraphics[width=\columnwidth]{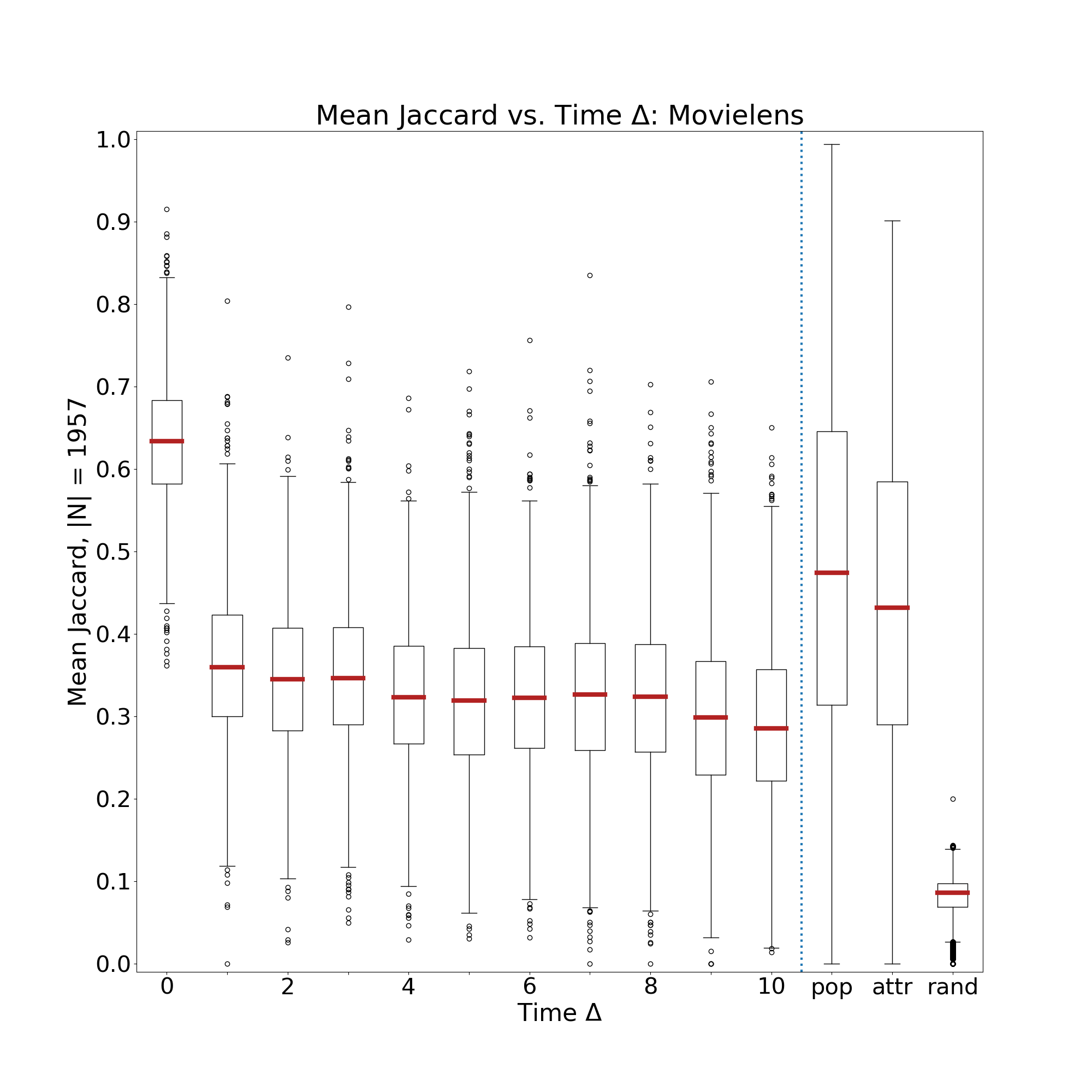}
  \caption{Predictability distributions vs Time $\Delta$} 
  \label{subfig:deltas_movielens}
\end{subfigure}
\begin{subfigure}{.33\textwidth}
\centering
 \includegraphics[width=\columnwidth]{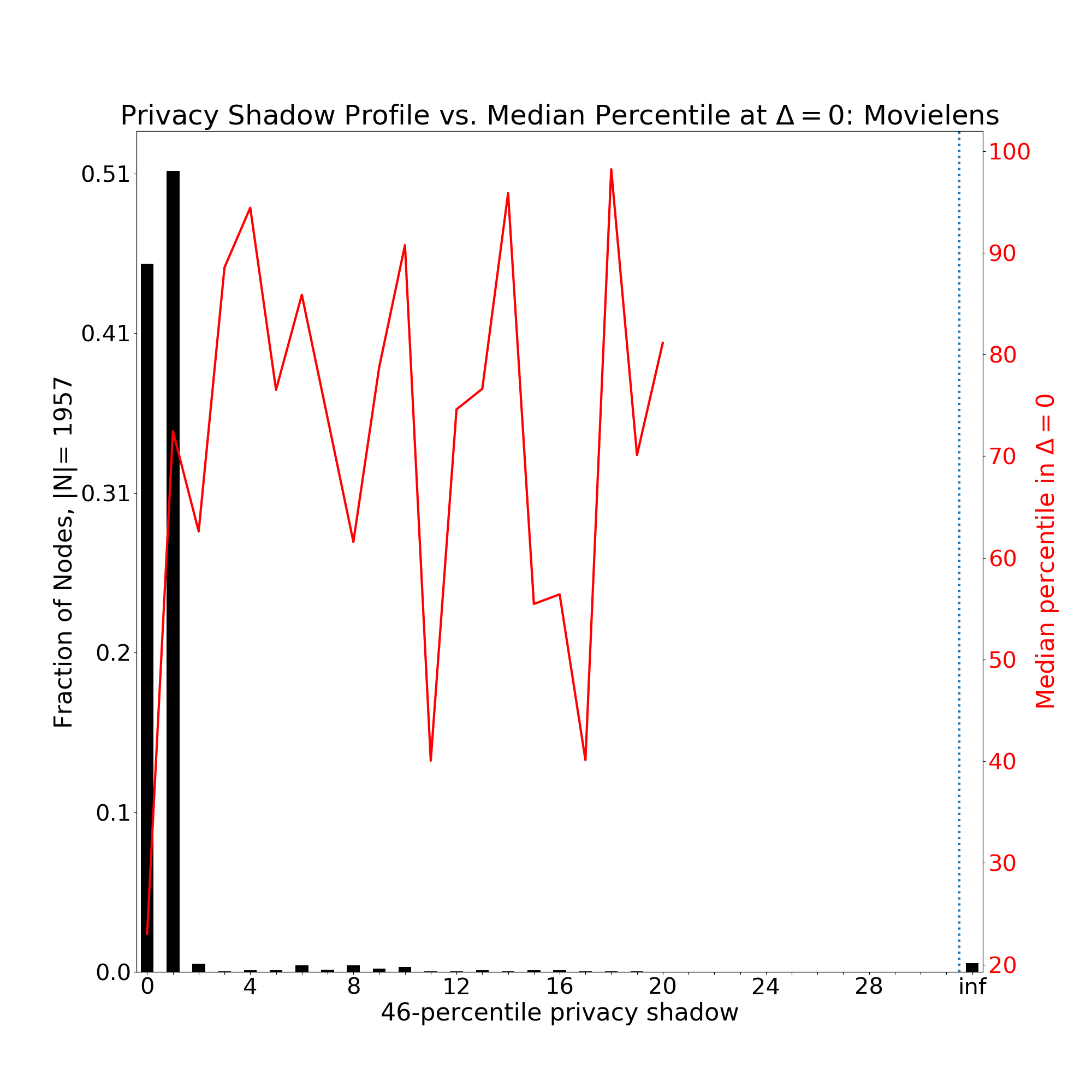}
 \caption{Privacy Shadow Profile vs. Median Score}
 \label{subfig:movielens_bounded} 
\end{subfigure}
\begin{subfigure}{.33\textwidth}
\centering
  \includegraphics[width=\figwidth\columnwidth]{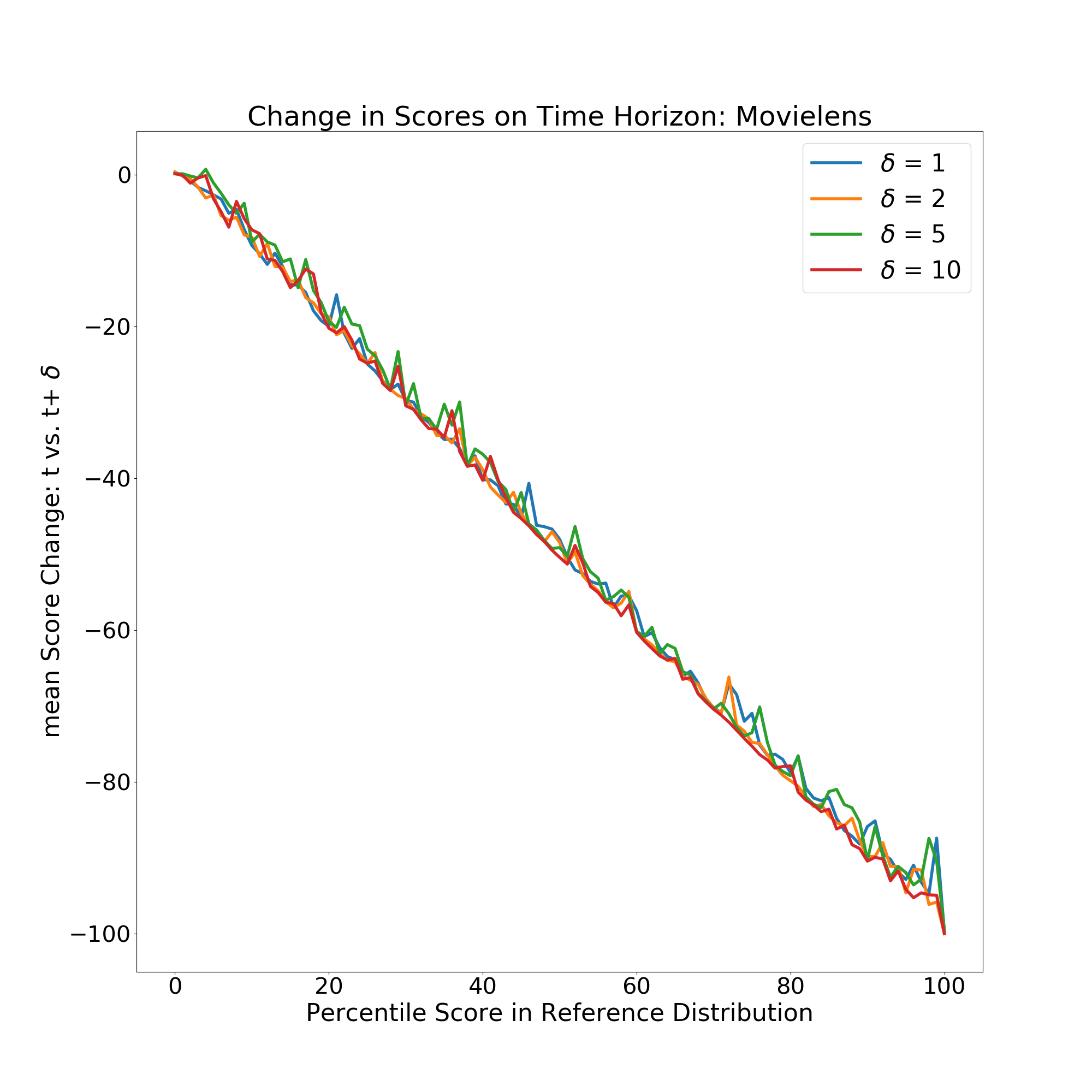}
  \caption{Score vs. Time-lagged Score} 
  \label{subfig:movielens_meandiff}
\end{subfigure}
\caption{Results on \ml{}: This dataset shows very weak network signal. The predictive performance at time $\Delta$=0 (a) is much higher than non-network baselines but reduces sharply in the next timestep. (b) shows that 46 percent of nodes are no better than the non-network baseline and 96\% of nodes have a privacy shadow of length 0 or 1. Finally (c) confirms that even at $\delta$=1 is below the baseline distribution at any predictability.}
\label{fig:movielens_all}
\end{figure*}

In this section we qualitatively compare empirical results across datasets and summarize our measurement and prediction results on both \ml{} and \aminer{} datasets. 

Figure \ref{fig:movielens_all} summarizes the evaluation of the \ml{} dataset. Overall, the induced $k$-NN no more predictive than baselines at \textit{any} time horizon. We verified this for a range of $k$ values and training window lengths $w$. Also, the size of the cohort is much smaller than other datasets, indicating there are fewer regular, long-term users in \ml{}. 

Figure \ref{subfig:deltas_movielens} shows that at expectation, the network performance drops by $0.3$ after one time step, where both of the population baselines outperform the network. The privacy shadow profile in Figure \ref{subfig:movielens_bounded} shows the network model is not predictive for nearly all nodes after one time step. Finally, Figure \ref{subfig:movielens_meandiff} follows the lower bound along $y=-x$. This means that the network model performs at the very bottom of the reference distribution for all subsequent timesteps $\delta$. This is a spurious network definition which our measurement detects. This is likely due to feedback with a recommendation engine which may prompt users on similar films who have no long-term correlation, e.g. during cold start.\footnote{this explanation was provided by one of the collaborators on this dataset.}

\begin{figure*}[ht]
\centering
\begin{subfigure}{.33\textwidth}
  \includegraphics[width=\figwidth\columnwidth]{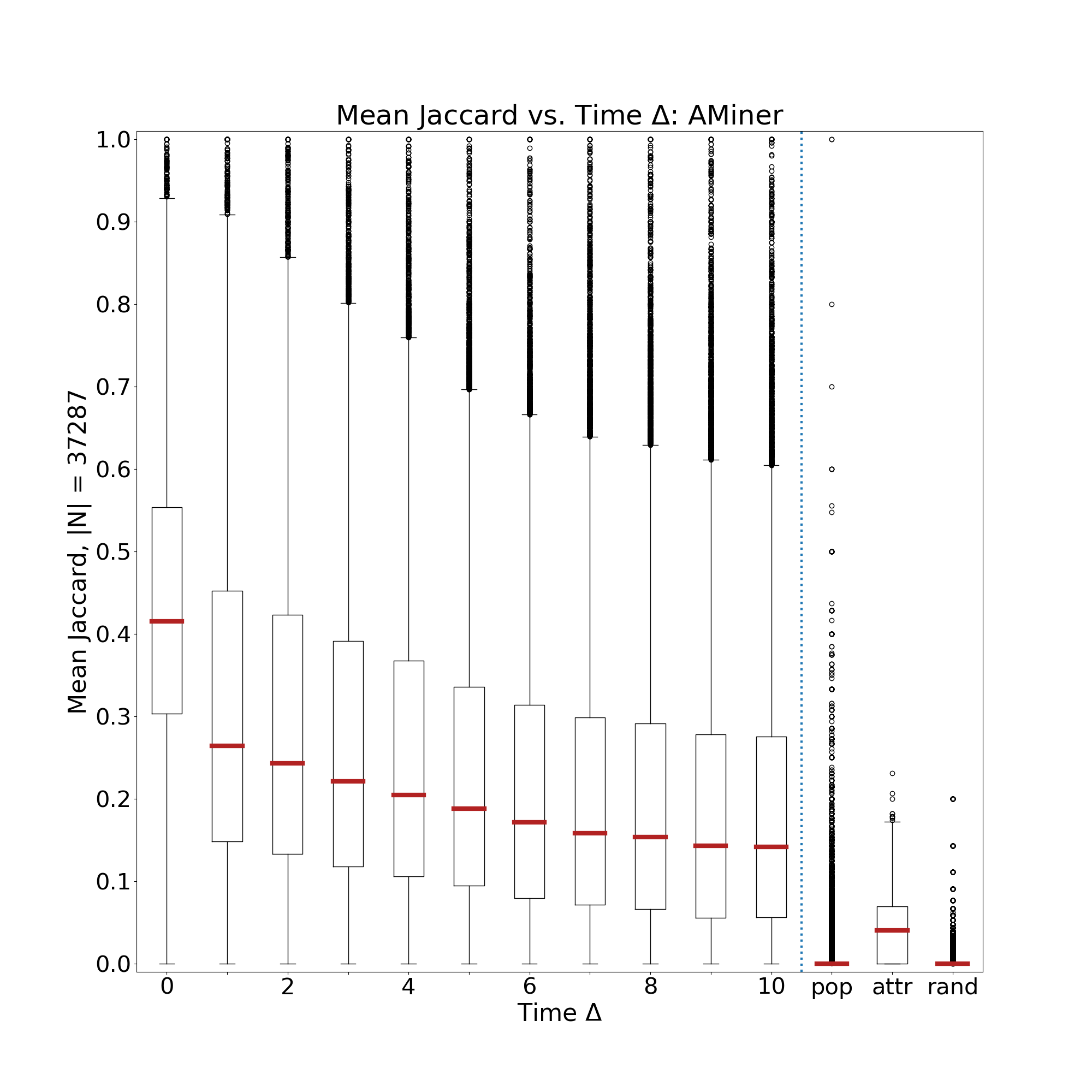}
  \caption{Predictability distributions vs Time $\Delta$} 
  \label{subfig:deltas_academic}
  \end{subfigure}
\begin{subfigure}{.33\textwidth}
  \includegraphics[width=\figwidth\columnwidth]{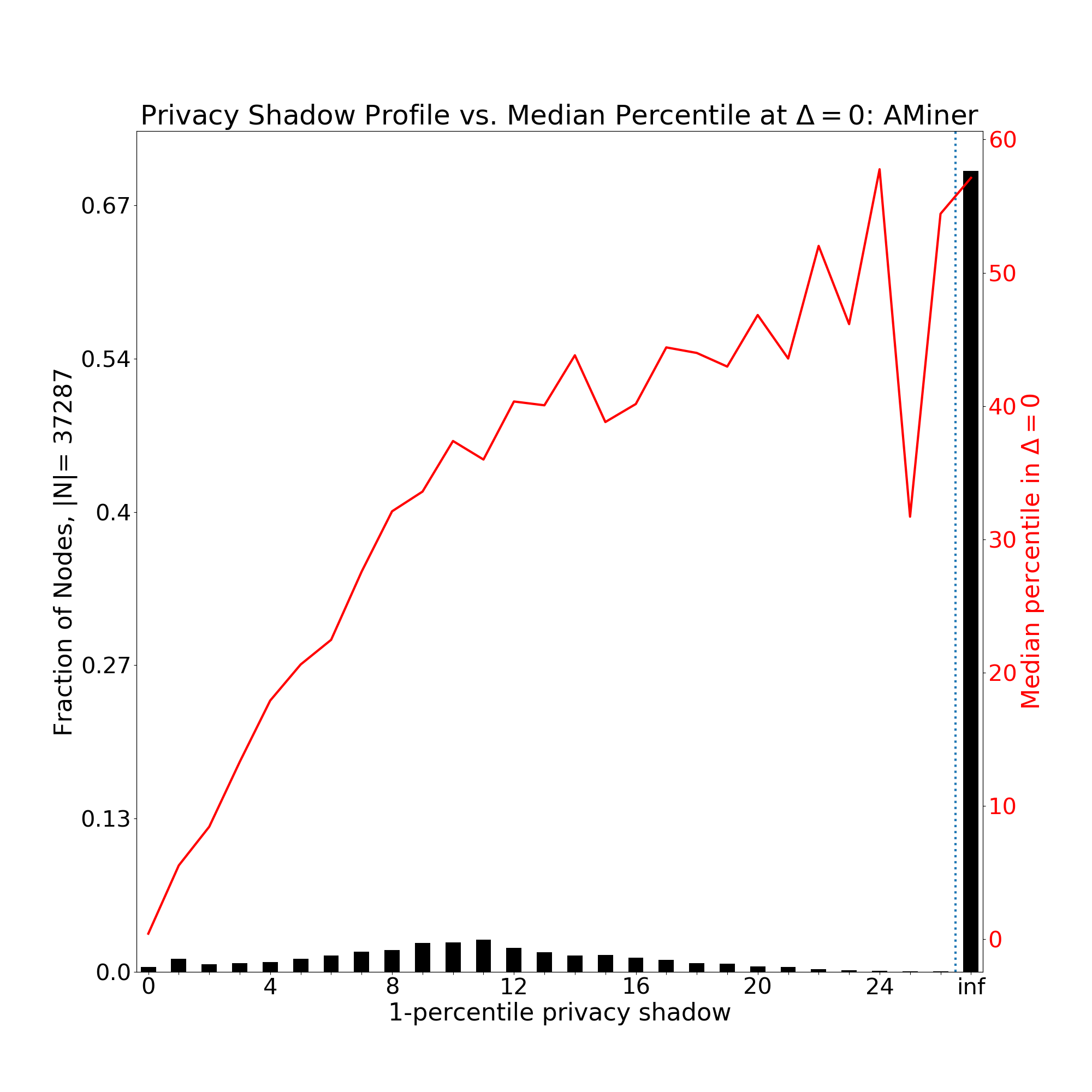}
  \caption{Privacy Shadow Profile vs. Median Score} 
  \label{subfig:aminer_bounded}
  \end{subfigure}
\begin{subfigure}{.33\textwidth}
\centering
  \includegraphics[width=\figwidth\columnwidth]{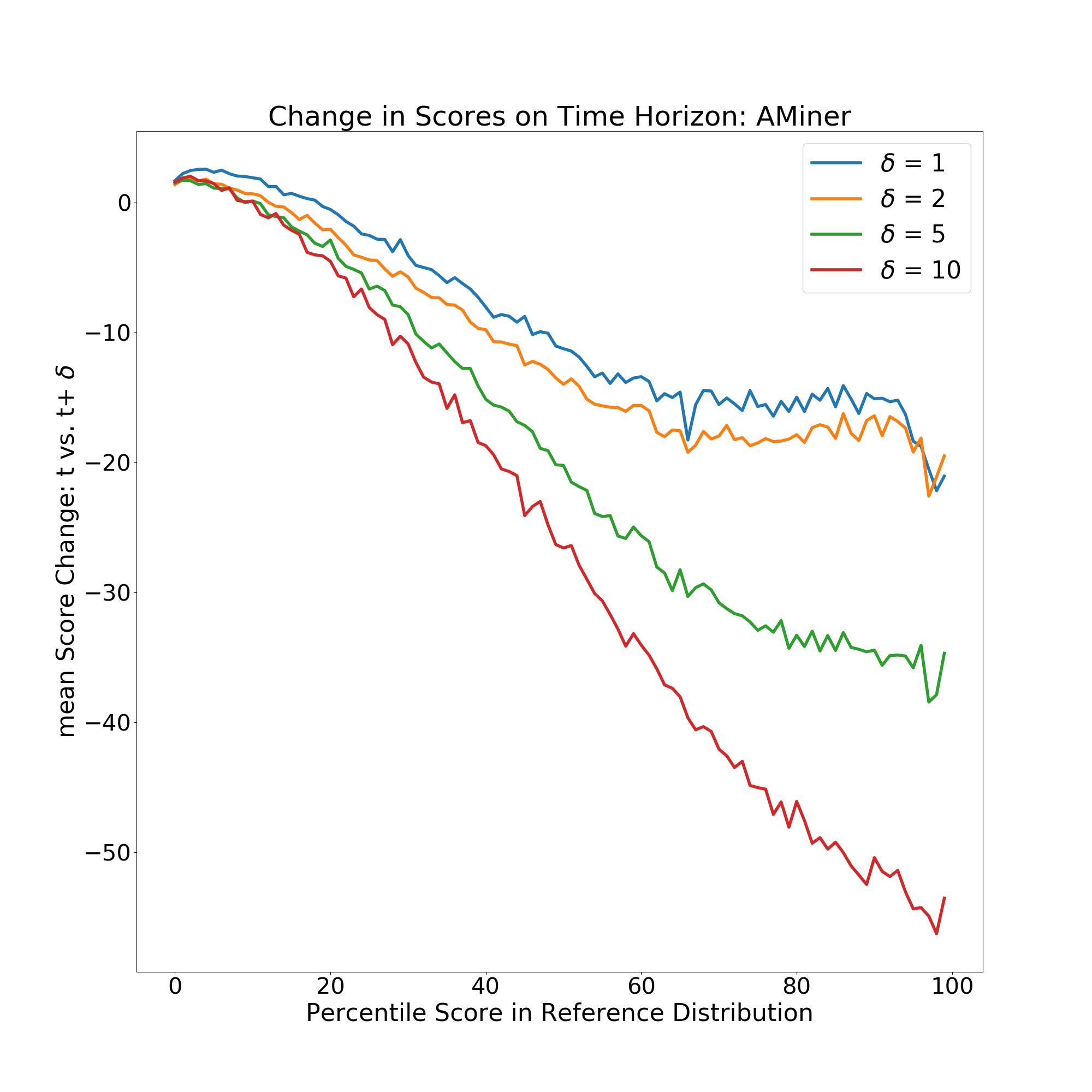}
  \caption{Score vs. Time-Horizon Score} 
  \label{subfig:aminer_meandiff}
\end{subfigure}
\caption{Results on \aminer{}. In (a) the co-authorship network is very predictive compared to the non-network baseline which fails for both node and attribute sampling. (b) Nearly 75\% of nodes are indefinitely predictable by a node's \textit{single-year} co-authorship neighborhood. (c) a node's degradation over varying time-horizons. For highly predictive nodes > 90 percentile, the network model is still greater than the 30-th percentile of the reference distribution after \textit{10 years}.}
\label{fig:aminer_all}
\end{figure*}

Figure \ref{fig:aminer_all} summarizes results on the \aminer{} dataset. The network in \aminer{} is co-authorship network on annual time-steps. Unlike \lfm{} and \ml{} datasets, we use one year as the training window. Recall that we predict the distribution a node's publication keywords over time. 

Figure \ref{subfig:deltas_academic} shows that the network model re-infers approximately 40\% at $\Delta=0$ in expectation. This is a notable result since the unique keywords are several orders of magnitude larger than the prior two datasets (634K unique keywords). Due to the specialization of labels within the network, all of the baselines perform very poorly. This yields a very poor privacy shadow threshold at $1$\%. It is very important that our privacy shadow threshold could greatly increase with a better zero-knowledge baseline. Providing this model would be complimentary to our methodology and we could measure the \aminer{} dataset with greater fidelity.

Figure \ref{subfig:aminer_bounded} agrees with the previous result, demonstrating that with respect to the baseline, the network model is indefinitely predictive for most nodes. 

Finally, Figure \ref{subfig:aminer_meandiff} shows that unlike \lfm{}, the \aminer{} dataset does not linearly scale with the percentile (x-axis). Instead, high-ranked observations ($> 60$ percentile) degrade more slowly  $\delta=1,2,5$ (years), but \textit{not} at $\delta=10$. This may be an indication of the time scale stable collaborations within this domain.

\begin{figure*}
\centering
\begin{subfigure}{.49\textwidth}
\centering
  \includegraphics[width=\figwidth\columnwidth]{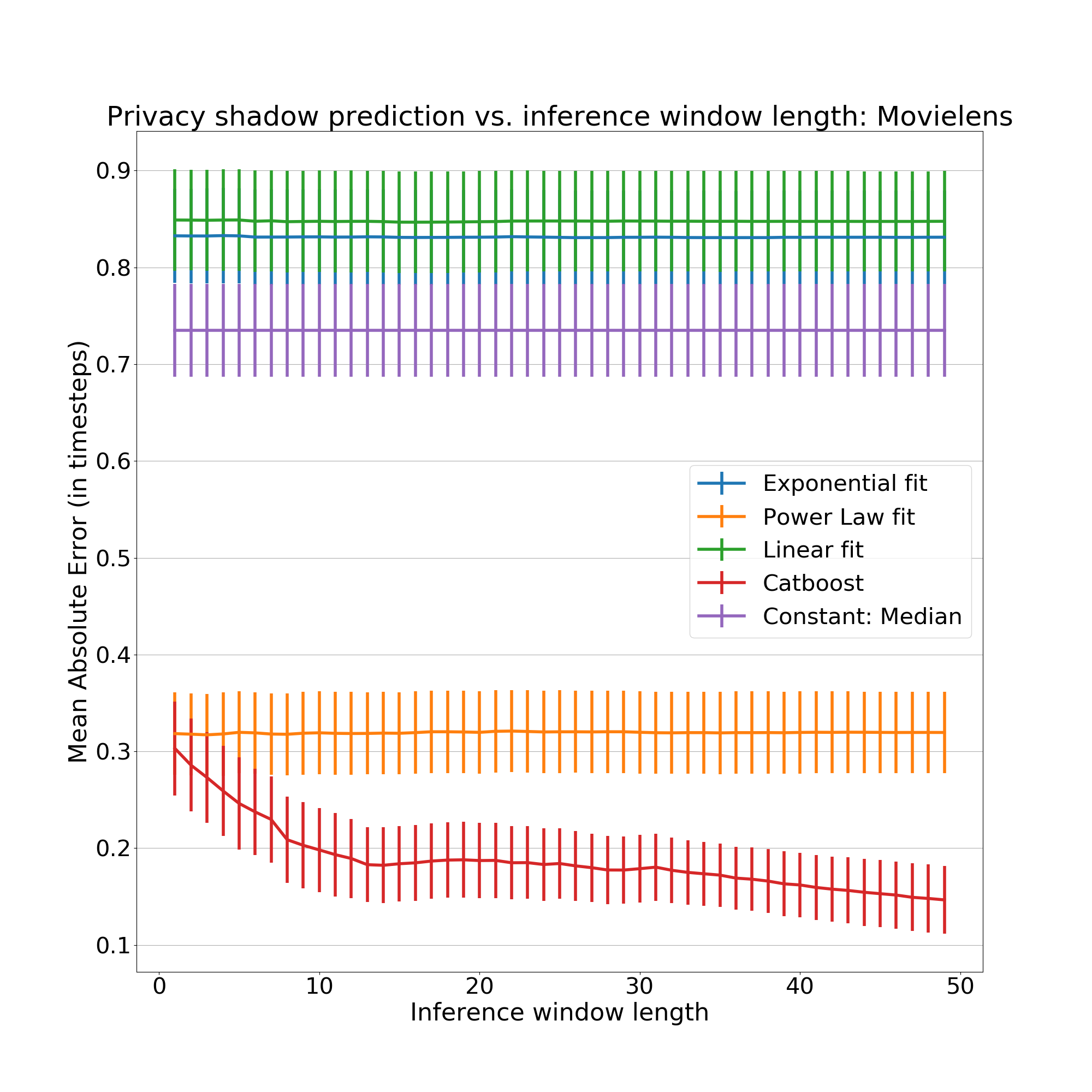}
  \caption{\ml{} privacy shadow length prediction} 
  \label{subfig:movielens_shadow_pred}
\end{subfigure}
\begin{subfigure}{.49\textwidth}
\centering
  \includegraphics[width=\figwidth\columnwidth]{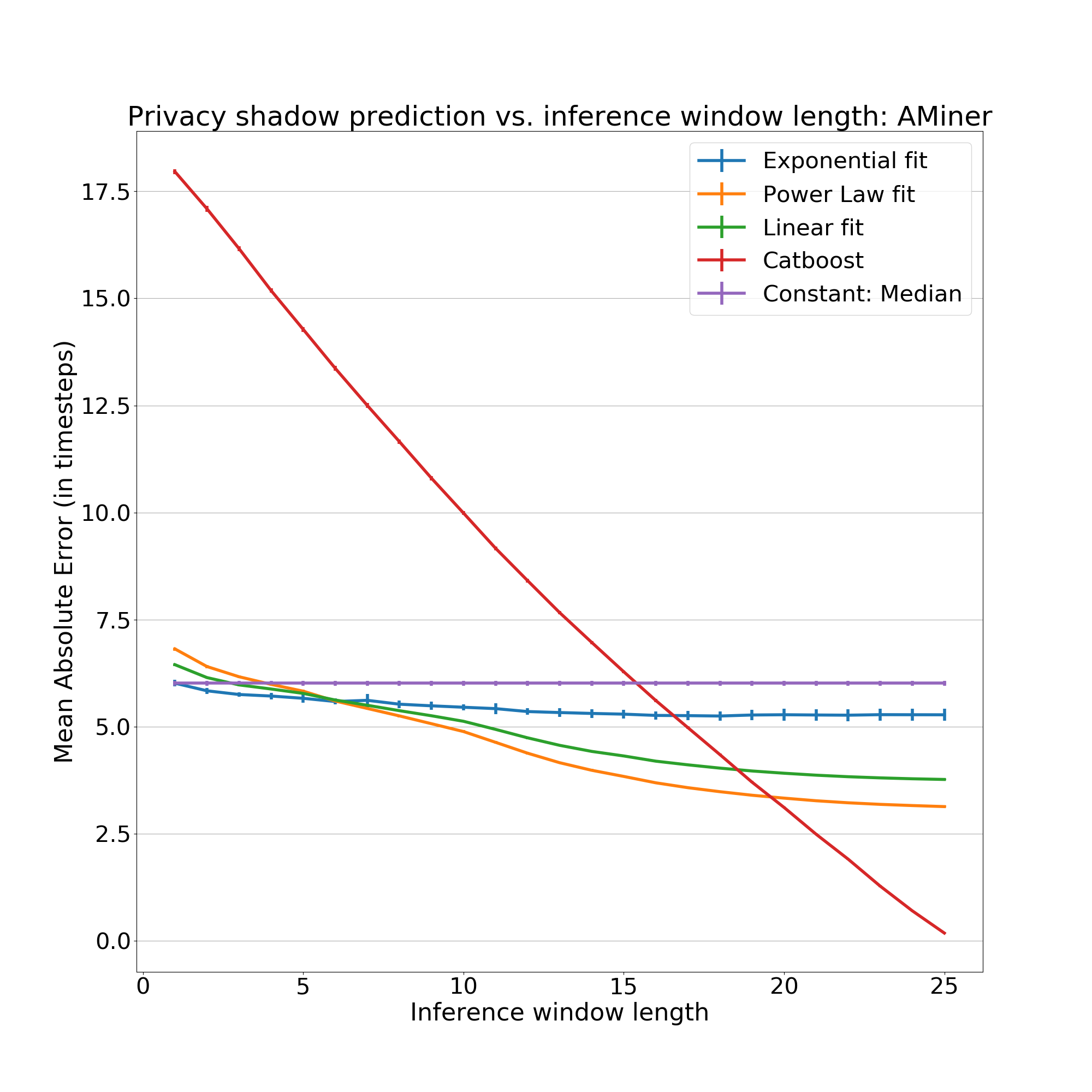}
  \caption{\aminer{} privacy shadow length prediction} 
  \label{subfig:aminer_shadow_pred}
\end{subfigure}
\caption{Privacy Shadow length prediction for \ml{} and \aminer{} datasets. In each figure, the length of the privacy shadow is predicted at model inference, varying the data available to the model by yielding null data for $t > x$}
\label{fig:movielens_aminer_predictions}
\end{figure*}

%  \begin{table}[h]
%  	\centering
%         \begin{tabular}{c||ccc}
%         Dataset$\backslash{}$ $y=ax^\lambda +c$ & a & $\lambda$ & c \\
%         \hline \hline
%         Last.fm    & $-17.65 \pm 15.75$ & $0.42 \pm 0.88$  & $46.47 \pm 29.40$   \\
%         Movielens  & $-47.74 \pm 28.43$   & $0.02 \pm 0.14$   & $49.28 \pm 28.91$   \\
%         AMiner     & $-21.13 \pm 20.39$   & $0.47 \pm 1.03$   & $48.92 \pm 30.57$  
%         \end{tabular}
%  	\caption{Mean and standard deviation of selected fit parameters for individual node score trajectories. For a node in the test set, we select the fit parameters of it's nearest neighbor in training, given a sequence length observed in test.} 
%  	\label{tab:line_fit_params}
%  \end{table}

%  \begin{table}[h]
%  	\centering
%         \begin{tabular}{c||cc}
%         Dataset$\backslash{}$ $y = ax + b$ & a & $b$ \\
%         \hline \hline
%         Last.fm    & $-0.29 \pm 0.47$ & $25.06 \pm 22.78$  \\
%         Movielens  & $-0.12 \pm 0.10$   & $4.66 \pm 3.42$   \\
%         AMiner     & $-0.75 \pm 1.27$   & $27.82 \pm 25.07$   
%         \end{tabular}
%  	\caption{Mean and standard deviation of selected fit parameters for individual node score trajectories. For a node in the test set, we select the fit parameters of it's nearest neighbor in training, given a sequence length observed in test.} 
%  	\label{tab:line_fit_params}
%  \end{table}

Figure \ref{fig:movielens_aminer_predictions} shows results for predicting the privacy shadow from rank trajectories. In Figure \ref{subfig:movielens_shadow_pred} for \ml{}, we see that all models perform trivially well. In this instance, the median is a constant model yielding $0$. We are less concerned with the relative performance of the models. Instead, it again validates that the \ml{} network is simply not predictive over time.

Figure \ref{fig:movielens_aminer_predictions} shows these results on \aminer. While the privacy shadow profile (Figure \ref{subfig:aminer_bounded}) looks very easy for a predictive model to learn, it does surprisingly poorly with respect to the baselines. We represent infinity as $T+1$, where T is the time series length. The median model yields $T + 1$. In practice, we therefore predict at $5$ \textit{years} MAE after $15$ \textit{years} of prediction trajectory. A better baseline model would yield a more informative label distribution to learn on. We will investigate this in future work. 

\section{Conclusions}

In this work, we proposed the \textbf{privacy shadow}, a simple methodology to measure node predictability over time from its prior local neighborhoods. The privacy shadow measures nodes over arbitrary timesteps against non-network, zero-information methods. We demonstrate that the length of the privacy shadow can be predicted from the simple rank trajectories from our methodology. Finally, we argue this measure can yield better guidance for users regarding the secondary effects of network structure on their privacy, even without direct access to their data. In this initial work, we motivated the privacy shadow and demonstrated measuring it empirically. 

Future work is in several important directions. First, a better understanding of privacy shadow dynamics under well-understood synthetic random graphs and parametric attribute distribution assumptions could provide better confidence bounds for prediction.  

Similarly, dynamic network models provide powerful randomization methods to better parameterize and infer the rates of rank degradation, e.g. as a function of transitioning to random models.  

\bibliographystyle{ACM-Reference-Format}
\bibliography{sample-base}

\appendix

\balance
\end{document}